\documentclass[12pt]{article}

\newcommand{\al}{\alpha} 
 
\newcommand{\als}{\frac{\alpha_s}{\pi}}
\newcommand{\alsb}{\left(\frac{\alpha_s}{\pi}\right)}

\newcommand{\ice}[1]{\relax}
\newcommand{\be}{\begin{equation}}
\newcommand{\ee}{\end{equation}}
\newcommand{\ba}{\begin{eqnarray}}
\newcommand{\ea}{\end{eqnarray}}
\newcommand{\MSsch}{\overline{\rm MS}}
\newcommand{\mts}{M_\tau^2}
\begin{document} 
\begin{flushright}
{\bf MZ-TH/99-46}\\
\end{flushright}
\vspace*{0.4cm}
\begin{center} 
{\Large \bf Asymptotic structure of perturbative series
for $\tau$ lepton decay observables:
$m_s^2$ corrections
}

\vspace{0.6truecm}

{\large J.G.~K\"orner$^1$, F.Krajewski$^1$,
A.A.~Pivovarov$^{1,2}$}\\[.1cm]
$^1$ Institut f\"ur Physik, Johannes-Gutenberg-Universit\"at,\\[-.1truecm]
  Staudinger Weg 7, D-55099 Mainz, Germany\\[.2truecm]
$^2$ Institute for Nuclear Research of the\\[-.1truecm]
  Russian Academy of Sciences, Moscow 117312
\vspace{0.6truecm}
\end{center}

\vskip 1cm
\centerline {\bf Abstract}
\noindent 
In a previous paper \cite{one} we performed an analysis of 
asymptotic structure of perturbation theory series for
semileptonic $\tau$-lepton decays in massless limit.
We extend our analysis
to the Cabibbo suppressed $\Delta S=1$
decay modes of the $\tau$ lepton. In particular 
we address the problem of $m_s^2$ corrections
to theoretical formulas. The 
properties of 
the asymptotic behavior of the finite order perturbation theory series
for the coefficient functions of the $m_s^2$ corrections
are studied.
\thispagestyle{empty}
\newpage
\section{Introduction}
The accuracy of 
experimental
data for $\tau$ lepton decays makes it feasible now 
to extract the spectral density of the Cabibbo suppressed $\Delta S =1$
decay modes through detecting strange hadrons
and even to pin down the
tiny difference with 
the Cabibbo favored $\Delta S =0$
case due to the non-vanishing strange quark mass
\cite{exp,PDG}.
One of the main problems in obtaining
precise theoretical formulas is 
the strict control over the convergence of perturbation theory
(PT) series 
and the error due to its truncation
\cite{one}.
For the nonstrange decay channels (Cabibbo favored) 
this problem is now an actual problem --
the theoretical uncertainty has already reached a limiting value existing 
due to asymptotic nature of the PT series. This value is comparable in
magnitude
with the experimental error.
For the Cabibbo suppressed modes the 
experimental errors are still larger than theoretical uncertainties.
However, with the accuracy
of experimental data permanently improving
the limiting theoretical precision within FOPT 
is becoming a major problem of theoretical analysis in general and 
of the extraction 
of the strange quark mass $m_s$ from $m_s^2$
corrections in particular.

From the theoretical point of view 
one of the central quantities of interest for the Cabibbo suppressed modes 
from theoretical point of view is 
the correction to hadronic spectral density arising from
the nonvanishing 
$s$-quark mass. 
This makes the description different compared to 
the massless (Cabibbo favored 
or $ud$) case.
The $m_s^2$ corrections to the spectral densities
have been calculated with a high
degree of accuracy within perturbation 
theory in the strong coupling constant (e.g. \cite{msNPB}). 

In the present note we determine 
the ultimate
theoretical precision reachable for $m_s^2$ corrections
within a finite order
perturbation theory analysis.
We follow closely the lines of ref. \cite{one}
and reach 
our conclusions in a renormalization scheme invariant way.

The basic observable is 
the normalized $\tau$ lepton decay rate into hadrons
written in the standard form
\be 
  \label{rate}
R_\tau={\Gamma(\tau \rightarrow h\nu)\over 
\Gamma(\tau \rightarrow l\nu\bar\nu)}
=N_c S_{EW}((|V_{ud}|^2(1+\delta_{ud})+|V_{us}|^2(1+\delta_{us})) \, .
\ee
The leading terms in eq.~(\ref{rate}) are the parton model 
results while the terms $\delta_{ud}$ and $\delta_{us}$ 
represent the effects of QCD interaction
and (in case of nonvanishing quark masses) mass effects
\cite{SchTra84,Bra88,Bra89,NarPic88,BraNarPic92}.
$V_{ud}$ and $V_{us}$ are matrix elements of the weak mixing matrix and
$S_{EW}$ describes the electroweak radiative corrections to the
$\tau$-decay rate. 

In general, hadronic 
observables in the $\tau$ system are related to the two-point correlator
of
hadronic currents with well established and simple analytic properties
-- this makes the comparison of experimental data with
theoretical calculations very clean. This feature makes
$\tau$ lepton physics an important area of particle phenomenology 
where theory (QCD) can be confronted with experiment 
to a level of very high precision.

The correlator (here 
we concentrate only on strange hadronic current, i.e.
the term proportional to $V_{us}$)
has the form
\begin{equation}
\Pi_{\mu\nu}(q) = 12 \pi^2 i \int dx e^{iqx}
\langle T j_{\mu}(x) j_{\nu}^{\dagger} (0) \rangle
=  q_{\mu}q_{\nu} \Pi_q(q^2)+g_{\mu\nu}  \Pi_g(q^2)   
\label{correlator}
\end{equation}
with $j_{\mu}(x) = \bar{u}\gamma_{\mu}(1-\gamma_5) s$.
$\Pi_q(q^2)$ and $\Pi_g(q^2)$ are invariant scalar functions. 
We work
within QCD with three light quarks and do 
not consider corrections due to heavy quarks ($c$-quark) that 
would enter
in higher orders of PT through loop effects \cite{chet93}.
The correlator is normalized to unity in the leading parton model approximation
with massless quarks.

The theoretical expression for the QCD part of the decay
rate into strange hadrons is given by 
(with $N_c |V_{us}|^2  S_{EW}$ factored out) 
\begin{equation}
  \label{int-qg}
R_{\tau} =\int_0^{M_\tau^2}2 \left(1-{s\over M_\tau^2}
\right)^2 \left(R_q(s)-
\frac{2}{M_\tau^2}R_g(s) \right)
{ds\over M_\tau^2} 
\end{equation}
with $R_q(s)$ and $R_g(s)$
being the absorptive parts of the structure functions 
$\Pi_q(q^2)$ and $\Pi_g(q^2)$.
The masses of the light quarks $(u,d)$ can be neglected.
We study $m_s^2$ corrections for the Cabibbo suppressed decay modes. 
The representation of the total decay
rate in terms of the absorptive parts of the structure functions 
$\Pi_q(q^2)$ and $\Pi_g(q^2)$ is convenient from the point of view of
their analytic properties in the complex $q^2$-plane.
The physical decomposition of the correlator (\ref{correlator}) reads
\be
\label{spindecomp}
\Pi_{\mu\nu}(q) = (q_{\mu}q_{\nu} -q^2 g_{\mu\nu})\Pi_T(q^2)
+ q_{\mu}q_{\nu} \Pi_L(q^2)
\ee
where the $\Pi_T(q^2)$ part contains only spin 1 contributions
and $\Pi_L(q^2)$ contains only spin 0 contributions.
The relation between the two sets of invariant functions 
$\Pi_{T,L}(q^2)$ and $\Pi_{q,g}(q^2)$
describing the
correlator eq.~(\ref{correlator})
reads
\be \label{physdec}
\Pi_T(q^2)=\frac{\Pi_g(q^2)}{-q^2}, \quad
\Pi_L(q^2)=\Pi_q(q^2)+\frac{\Pi_g(q^2)}{q^2} \ .
\ee
In terms of the physical (definite spin) invariant functions 
eq.~(\ref{int-qg}) reads    
\begin{equation}
  \label{int-TL}
R_{\tau}=\int_0^{M_\tau^2}2 \left(1-{s\over M_\tau^2}
\right)^2
 \left( \left( 1+2{s\over M_\tau^2} \right) R_T(s)+R_L(s) \right)
{ds\over M_\tau^2} \ .
\end{equation}
The longitudinal part of the spectral density $R_L(s)$ vanishes if all
quarks are assumed to be massless.
On expanding $\Pi_q(q^2)$ and $\Pi_g(q^2)$ 
in $m_s^2/q^2$ and keeping only the leading term in this expansion
one has
\be
\label{expmq}
\Pi_q(q^2)=\Pi(q^2)+3\frac{m_s^2}{q^2}\Pi_{mq}(q^2)
\ee
\be
\label{expmg}
\Pi_{g}(q^2)=-q^2 \Pi(q^2) + \frac{3}{2} m_s^2   \Pi_{mg}(q^2)
\ee
where $\Pi(q^2)$ 
is the invariant function already known from the mass zero 
case.
The functions $\Pi_\#(Q^2)$ with $Q^2=-q^2$ are computable in
perturbation theory in the deep Euclidean region $Q^2\rightarrow \infty$.  
The results of the PT calculation read
\begin{eqnarray} 
\label{fullinf}
-Q^2\frac{d}{dQ^2}\Pi(Q^2)\Big|_{Q^2=\mu^2}
&=&1 + \als + k_1 \left( \als \right)^2 + k_2 \left(\als \right)^3 
+k_3 \left( \als \right)^4 + O(\alpha_s^5) \nonumber \\
-Q^2\frac{d}{dQ^2}\Pi_{mg}(Q^2)\Big|_{Q^2 =\mu^2}
&=&1+\frac{5}{3}\als + k_{g1}\left(\als \right)^2 
+ k_{g2}  \left( \als \right)^3 + k_{g3} 
\left(\als \right)^4 + O(\alpha_s^5)\nonumber \\
\Pi_{mq}(Q^2)\Big|_{Q^2=\mu^2}
&=& 1 + \frac{7}{3} \als + k_{q1}  \left( \als \right)^2
+ k_{q2}  \left( \als \right)^3 + O(\alpha_s^4)
\end{eqnarray}
Even though the fourth order 
$\overline{\rm MS}$ scheme coefficient $k_3$ and the coefficients
$k_{q2},\;k_{g3}$ are not known at present we retain their 
contributions since we want to dispose on them as
free parameters for later considerations.
The numerical values of the other coefficients in 
the $\overline{\rm MS}$ scheme are given in the Appendix:
these are needed as reference numbers for our transformations 
to other 
more appropriate schemes. 
The running coupling and mass are renormalized at the scale 
$\mu$. The light quarks $u$, $d$ are taken to be massless. 
Eq.~(\ref{fullinf})
constitutes 
the complete theoretical information 
necessary for our fixed order perturbation theory analysis of
$m_s^2$ corrections.
The corresponding expressions 
with an explicit $Q^2$ dependence can be found by inserting 
the expansion of the running coupling constant 
\ba \label{runcoupl}
\frac{\alpha_s(Q^2)}{\pi} &=& \als + \beta_0 L (\als)^2 + (\beta_1 L +
\beta_0^2 L^2) (\als)^3
+ (\beta_2 L +\frac{5}{2}\beta_1\beta_0 L^2 + \beta_0^3 L^3) (\als)^4
\nonumber \\
 && + (\beta_3 L + 3 \beta_0 \beta_2 L^2 + \frac{3}{2} \beta_1^2 L^2 + 
\frac{13}{3} \beta_0^2 \beta_1 L^3 + \beta_0^4 L^4) (\als)^5+ \ldots 
\ea
and that of the running mass
\ba 
\label{runmass}
\frac{m_{s}(Q^2)}{m_s(\mu^2)} &=& 1 + L  \gamma_0 (\als) + (\frac{1}{2} L^2
\beta_0 \gamma_0 + \frac{1}{2} L^2 \gamma_0^2 + L \gamma_1)(\als)^2 + ...
\ea
Here $\beta_i$ and $\gamma_i$ are the appropriate 
coefficients of $\beta$- and  
$\gamma$-functions 
describing the evolution (running) of the coupling 
and mass and 
\be
L= \ln(\frac{\mu^2}{Q^2}).
\ee
The coupling constant $\alpha_s$ in eq. (\ref{runcoupl},
\ref{runmass})
in taken at a genuine normalization point $\mu$.
In the present note we do not systematically 
discuss non-perturbative effects stemming 
from standard power corrections \cite{SVZ}.
The standard power corrections arise from 
nonvanishing vacuum expectation
values of local operators 
within the operator product expansion 
and are relatively small. They can be simply
accounted for if necessary. 
They do not mix with the $m_s^2$ corrections.
The coefficient
functions of the local operators are 
known in low orders of the perturbative expansion
and 
there is no necessity for a thorough analysis of their convergence
properties at present.

\section{Natural strange quark masses for 
internal \\
perturbation theory description of $m_s^2$ corrections}
We restrict our attention only to the new features that appear due to 
the mass
corrections when compared to our previous analysis
\cite{one}.
The appropriate quantities to consider
are moments of a spectral density
\begin{equation}
  \label{intmom}
M_n=(n+1)\int_0^1 \rho(x) x^n dx \ .
\end{equation}
We often use $M_\tau$ as a unit of mass which leads to dimensionless
variable $x=s/M_\tau^2$.
Note that within finite order
perturbation theory the moments eq.~(\ref{intmom})
coincide with the results of a contour integration
\cite{cont,cont1,cont2,Pivtau}
because of analytic properties of the functions $\ln^p z$.
The moments of the hadronic
spectral densities are internal characteristics of the hadronic decays of
the $\tau$ system 
and it is instructive to describe these moments in internal variables.
In the massless limit within FOPT
there is only one independent internal variable -- 
the effective coupling $a(s)$ which is defined directly on the physical cut 
through the relation
\begin{equation}
  \label{defofa}
\rho(s)=1+a(s) 
\end{equation}
and studied in ref.~\cite{one}.
All the constants that may appear due to a particular
choice of the renormalization 
scheme are absorbed into the definition of the effective charge e.g.
\cite{effsch,ksch,kksch,effDh}.
In our present analysis 
the effective charge is determined by the massless piece $\Pi(q^2)$
of the correlator (\ref{correlator}) 
(see, eq. (\ref{expmq},\ref{expmg})).
Eq.~(\ref{defofa}) fixes the definition of the 
effective charge which is later used as an expansion parameter for
the mass corrections. 
With such a definition one retains consistency in the description
of the massless approximation for strange and non-strange modes.

The running of the coupling $a(s)$ defined in eq.~(\ref{defofa})
contains logarithms of $s$ with coefficients given by an effective
$\beta$-function
\be
\bar \beta(a)=s\frac{d a(s)}{ds}=-\beta_0 a^2
-\beta_1 a^3 -\bar \beta_2 a^4-\bar \beta_3 a^5 + O(a^6)
\ee
and reads
\begin{eqnarray}
\label{run}
a(s)&=&a + \beta_0 l a^2 + (\beta_1 l + \beta_0^2 l^2)
a^3+(\bar\beta_2 l+\frac{5}{2}\beta_1\beta_0 l^2 
+ \beta_0^3 l^3) a^4 \nonumber \\
&& + (\bar\beta_3 l + 3 \beta_0 \bar\beta_2 l^2 + \frac{3}{2}
\beta_1^2 l^2 + 
\frac{13}{3} \beta_0^2 \beta_1 l^3 + \beta_0^4 l^4)
a^5+ \ldots
\end{eqnarray}
where $a=a(M_\tau^2)$, $l=\ln(M_\tau^2/s)$.
Note that the expansion of $a(s)$ in 
eq.~(\ref{run}) has the same form as the one 
for $\alpha_s(s)/\pi$ but now $\bar\beta_2$ and $\bar\beta_3$ are 
coefficients of the effective $\beta$-function \cite{one} while 
$\beta_0$ and $\beta_1$ are renormalization group invariants.
The effective coupling $a$ can be expressed through the $\MSsch$
coupling constant
\be 
a = \als + 1.64 \alsb^2 -10.28 \alsb^3 + (-155.0 + k_3) \alsb^4 + ...
\ee
with $\alpha_s \equiv \alpha_s(M_\tau^2)$.
The massless spectral density reads \cite{one}
\begin{eqnarray}
\label{densold}
\rho(s)&=&1+a+2.25 a^2 l + a^3(4l + 5.063l^2)+ a^4(-25.7 l + 22.5 l^2 +
11.4 l^3) \nonumber \\
&&+a^5((-409.5 + 4.5 k_3)l - 149.4 l^2 + 87.75 l^3 + 25.63 l^4)+O(a^6) . 
\end{eqnarray}
At any fixed order of 
perturbation theory the effects of running die out for
the high order moments (large $n$ in eq.~(\ref{intmom}))
improving the convergence of the perturbation theory series. 
With the definition of the charge according to eq.~(\ref{defofa})
all high order corrections vanish as $n\rightarrow \infty$ 
for any 
fixed order of perturbation theory. 
However, for the mass corrections this is not true anymore
since
the strange quark mass introduces a new parameter.
In order to obtain only logarithms of energy 
in the spectral density one can redefine the $\MSsch$-scheme quark mass
appropriately
and absorb all remaining constants into the internal mass 
parameter.
Because there are two invariant functions that characterize the 
correlator (\ref{correlator}) the definitions of the
mass parameters 
may be
different for them. 
This redefinition is nothing but the change of subtraction scheme for
the given spectral density.
Note that the introduction of
a natural internal coupling parameter 
such as the effective charge $a(s)$
allows one to extend 
the perturbation theory series needed for the description of relations
between observables by one more term
as compared to the analysis in e.g. the $\MSsch$-scheme (e.g.
\cite{prl,renRS,brodsky1,brodsky}).
The same reason of obtaining one more term in the perturbation series 
is behind the definition of internal parameters
for the quark mass.

\subsection{The $g$-part of the correlator: function $\Pi_{mg}(q^2)$} 
First we consider the $g$-part of the correlator
which contains spin 1 contributions only.
We use the massless case
\cite{one} as a base for our analysis here.
We define a new mass parameter $m_g$ for the strange quark
to absorb all constants in the mass
correction of the spectral density into the new mass definition.
By definition
\ba \label{defofmg}
m_s^2(s)\rho_g^{\MSsch}(s;\al_s)\equiv
m_g^2(s;a)= m_g^2(\mts) \rho_g(s) .
\ea
The new mass parameter $m_g=m_g(\mts)$ is related to 
the $\MSsch$-scheme mass parameter $m_s=m_s(\mts)$
through
\ba 
\label{mg}
m_g^2&=&m_s^2 \left( 1+1.67 a - 5.87 a^2 - 51.0 a^3 +(-1342.5 -1.67 k_3 + k_{g3})
a^4+ O(a^5) \right)  \\
&=&m_s^2 \left( 1+1.67 \alsb  - 3.14 \alsb^2 - 87.4 \alsb^3 + \right. \nonumber \\ 
   &&  \left. ( -1750 + k_{g3} ) \alsb^4 + O \left(\alsb^5 \right) \right) \nonumber
\ea
The spectral density $\rho_g(s)$ for the $g$-part of the mass
correction contains only logarithms in fixed order PT expansion and reads
\begin{eqnarray}
\label{densg}
\rho_g(s)&=&1+2 a l + a^2 (8.05 l + 4.25 l^2) + a^3(5.3 l + 38.23
l^2 + 9.21 l^3 ) \nonumber \\
&&+a^4((-45.3  -2.0 k_3)l + 67.4  l^2 + 134.2 l^3 + 20.14l^4)+O(a^5) .
\end{eqnarray}
With the help of eq.~(\ref{mg}) the numerical value of 
$m_g^2$ can be determined in terms of the $\MSsch$-scheme
mass squared $m_s^2$ with an accuracy of about
7\% if eq.~(\ref{mg}) is evaluated up to third order. If the series is
(perturbatively) inverted, the $\MSsch$-scheme strange quark mass
can be extracted with an accuracy of about 4\% from the relation
\be
m_s^2 = m_g^2 (1 - 0.185 + 0.107 + 0.037 + (0.19 + 0.00025 k_3 - 0.00015
k_{g3})) .
\ee
This accuracy is sufficient at present for phenomenological
applications.

Given the expression for the spectral density eq.~(\ref{densg})
the whole analysis 
of ref.~\cite{one} applies.
The moments of the spectral density $\rho_g(s)$ in general behave worse than
in the massless case. 
This is understandable if one compares the coefficients of 
the logarithms
in the spectral densities (\ref{densold})
and (\ref{densg}). The coefficients are
larger overall in the mass correction case.
The basic objects one needs for the construction of observables are
moments of the spectral density $\rho_g(s)$
\begin{equation}
  \label{intmomg}
M_g(n)=(n+1)\int_0^1 \rho_g(s) s^n ds \ .
\end{equation}
We find
\begin{eqnarray}
\label{momff}
M_{g}(0)&=&1+ 2a + 16.6 a^2  +  137.0  a^3  
+ (1378.5-2.0 k_3)a^4  \nonumber \\
M_{g}(1)&=&1+ a + 6.15 a^2 + 28.67  a^3  
+ (141.97-1.0 k_3) a^4 \nonumber \\
M_{g}(2)&=&1+\frac{2}{3}a+3.63 a^2 + 12.31 a^3   
+(35.69 -0.67 k_3) a^4  \nonumber \\
M_{g}(3)&=&1+ \frac{1}{2}a +  2.54  a^2  + 6.97  a^3   
+ (11.58 -0.5  k_3 ) a^4   \nonumber \\
M_{g}(4)&=&1+ \frac{2}{5}a +  1.95  a^2  + 4.56  a^3   
+ (3.55 -0.4  k_3 ) a^4   \nonumber \\
&&\vdots \nonumber \\
M_{g}(100)&=&1+ \frac{2}{101} a + 0.081 a^2  + 0.060 a^3  
+ (-0.434 -0.0198 k_3) a^4 
\end{eqnarray}
Note that the unknown coefficient $k_{g3}$, which would appear in the 
fourth order coefficient of the moments in the $\MSsch$-scheme, is 
absorbed in the definition of the mass $m_g$.
Still the fourth order coefficient is not known because of its
dependence on $k_3$ which enters due to 
the charge redefinition. 
Indeed, this dependence has its origin in the definition of the mass $m_g$
(\ref{defofmg}).
The third coefficient of the effective $\gamma$-function
$\gamma_{g3}$
depends on $k_3$. This dependence affects the fourth order coefficient of
$\rho_g(s)$
through the running mass (\ref{runmass})
\be
\gamma_0 = 1, 
\quad \gamma_{g1} = 4.027,  
\quad \gamma_{g2} = 2.65,
\quad \gamma_{g3} = -22.65 - k_3 \, .
\ee
For large $n$ the moments behave better because the infra-red region 
of integration is suppressed.
Note that the coefficients of the series in 
eq.~(\ref{momff}) are saturated with the lowest power of logarithm
for large $n$ for a given order of perturbation theory, i.e. 
they are saturated with the highest coefficient of 
the effective $\beta$-function and $\gamma$-function.

It 
is instructive to compare the results in eq.~(\ref{momff})
with the $\MSsch$-scheme expansions given by 
\begin{eqnarray*}
M_{g}^{\MSsch}(0) &=&1+ 3.67 \als + 20.0 \alsb^2 + 110.1 \alsb^3
+(-256.3 + k_{g3})\alsb^4 \\
M_{g}^{\MSsch}(1) &=&1+ 2.67 \als + 6.32  \alsb^2  -38.98 \alsb^3
+(-1779 + k_{g3})\alsb^4 \\
M_{g}^{\MSsch}(2) &=&1+ 2.33 \als + 2.70  \alsb^2  -64.26  \alsb^3
+(-1865 + k_{g3})\alsb^4 \\
M_{g}^{\MSsch}(3) &=&1+ 2.17 \als + 1.06  \alsb^2  -73.18  \alsb^3
+(-1863 + k_{g3})\alsb^4 \\
M_{g}^{\MSsch}(4) &=&1+ 2.07 \als + 0.14   \alsb^2  -77.46  \alsb^3
+(-1851 + k_{g3})\alsb^4 \\
&&\vdots \\
M_{g}^{\MSsch}(100) &=&1+ 1.69 \als + 2.99 \alsb^2  -87.15  \alsb^3
+(-1755 + k_{g3})\alsb^4 \, .
\end{eqnarray*}
The advantage of the effective scheme against the 
$\MSsch$-scheme is apparent starting with 
$\al_s^3$ coefficient.
For moments larger than one ($n>1$)
in the $\MSsch$-scheme the series shows
already asymptotic 
growth in the third order while in 
the effective scheme the third coefficient
is still smaller
than the previous one. From its
construction it is clear that
the convergence behavior improves with higher moments in the
effective scheme while in $\MSsch$-scheme high moments 
become even worse than the lower ones
from the point of view of the structure of the perturbation theory
series. 
The anomalously small third order
coefficient
of the first moment in the $\MSsch$-scheme is the result of an accidental
cancellation 
of the contributions of logarithmic and constant terms because of the
particular choice of the scheme.
The general discussion of ref.~\cite{one} now applies.   
For our numerical estimates we take $a=0.111$ as
obtained
from the corresponding value of the $\MSsch$-scheme charge.
In the effective scheme the PT series for the moments read
\begin{eqnarray}
M_{g}(0) &=&1+ 0.222 + 0.204 + 0.187 + (0.21-0.0003 k_3) 
\nonumber \\
M_{g}(1) &=&1+ 0.111 + 0.076 + 0.039 + (0.022-0.00015 k_3)
\nonumber \\
M_{g}(2) &=&1+ 0.074 + 0.045 + 0.017 + (0.0054-0.00010 k_3)
\nonumber \\
M_{g}(3) &=&1+ 0.056 + 0.031 + 0.010 + (0.0018-0.000076 k_3)
\nonumber \\
M_{g}(4) &=&1+ 0.044 + 0.024 + 0.006 + (0.00054-0.000061 k_3) \, .
\end{eqnarray}
With the choice $k_3 = 100$ 
\cite{one}
the fourth order correction is smaller than the 
third term for the moments with $n<5$. A formal accuracy of about 0.7\%
can be obtained
if the zero order moment is excluded
using as an estimate the contribution of the smallest term.
With the standard Pad\'e estimate for $k_3 = 25 $ the moments become
\begin{eqnarray}
M_{g}(0) &=&1+ 0.222 + 0.204 + 0.187 + 0.20  \nonumber \\
M_{g}(1) &=&1+ 0.111 + 0.076 + 0.039 + 0.018 \nonumber \\
M_{g}(2) &=&1+ 0.074 + 0.045 + 0.017 + 0.003  \nonumber \\
M_{g}(3) &=&1+ 0.056 + 0.031 + 0.010 - 0.0001 \nonumber \\
M_{g}(4) &=&1+ 0.044 + 0.024 + 0.006 - 0.001 \, .
\end{eqnarray}
If one excludes $M_{g}(0)$ an accuracy of better than 2\% can be 
obtained. 

There is no value for $k_3$ which makes the forth order corrections 
of all moments smaller than the previous correction.
If $k_3$ has a value between $-46.5$ and $1.2$ all moments starting from
the first moment show no asymptotic growth in fourth order.  
Such a fine tuning of the unknown coefficient 
$k_3$ seems to be unrealistic. We thus conclude that asymptotic 
growth is unavoidable in fourth order in the $g$-part. The ultimate 
accuracy depends on the value of $k_3$ varying between 0.7\% and 
2\% if the zero order moment is excluded.
The invariant statement about the asymptotic
growth is that the system of moments $M_g(n)$ with $n=0$
included cannot be treated perturbatively at the fourth order 
of perturbation theory 
for the given numerical value of the expansion parameter 
$a=0.111$ if one wants to obtain an accuracy 
of the coefficient function in front of $m_s^2$ correction 
in $\Pi_g(q^2)$ amplitude better
than 15\% - 20\%.
This statement about the ultimate
accuracy of the set of moment observables
attainable in 
fourth order of perturbation theory is independent of whichever numerical 
value $k_3$ takes. 

The perturbation theory expansions for 
the system of moments with $(1-s)^n$ weight 
\be
\label{alsu}
\tilde M_g(n,0)= (n+1)\int_0^1 \rho_g(s) (1-s)^n ds =
(n+1)!\sum_{k=0}^n\frac{(-1)^k}{(k+1)!(n-k)!} M_g(k) \, 
\ee
show a much worse PT behavior. One has
\begin{eqnarray}
\label{altmomnum}
\tilde M_g(1,0)  &=& 1+0.333 + 0.332 + 0.336 
+ ( 0.397 - 0.00046 k_3)\nonumber \\ 
\tilde M_g(2,0)  &=& 1+0.407 + 0.429 + 0.461 
+ ( 0.569 - 0.00056 k_3) \nonumber \\
\tilde M_g(3,0)  &=& 1+0.463 + 0.509 + 0.572 
+ ( 0.728 - 0.00063 k_3) \, .
\end{eqnarray}
The
convergence of the series is obviously quite poor.
All moments $\tilde M_g(n,0)$ contain the contribution of 
$\tilde M_{g}(0,0)\equiv M_{g}(0)$ in the sum eq.~(\ref{alsu})
which by itself shows a bad behavior; the rest makes it worse.

To summarize,
the general PT structure of the moments for the 
$g$-part in the effective scheme with the new mass parameter $m_g^2$
is very similar to the massless 
part and there are no qualitatively new features found here
as compared to the analysis of the massless part in ref.~\cite{one}.

\subsection{The $q$-part of the correlator: function $\Pi_{mq}(q^2)$}
The $q$-amplitude $\Pi_q(q^2)$ contains contributions of
both spin 1 and spin 0 final states.
The correction to the $q$-part is different from the $g$-part
as concerns its analytic properties
in the $q^2$-plane -- it contains a $1/q^2$ singularity at the origin -- 
which necessitates a separate treatment. 
The explicit power singularity $1/q^2$ at the origin
of the function 
$\Pi_{mq}(q^2)/q^2$ makes the formulation
of the moments for the spectral density of the $q$-part
directly on the physical cut a bit tricky.
Indeed, because of
the $1/q^2$ factor the amplitude for the
$m_s^2$ correction
has no standard dispersion representation.
Rather the dispersion representation should be written as 
\be
\frac{\Pi_{mq}(Q^2)}{Q^2} = \int\frac{d\sigma(s)}{s+Q^2}
\ee
with a measure $d\sigma(s)$ which is not differentiable,
i.e. $d\sigma(s)\ne \sigma'(s)ds$ with some continuous $\sigma'(s)$.
However, it can be written in a more familiar form 
if a different weight is used
\be \label{pirhof}
\frac{\Pi_{mq}(Q^2)}{Q^2}=\int_0^\infty\frac{\rho_F(s)ds}{(s+Q^2)^2}=
-\frac{d}{d Q^2}F(Q^2)
\ee
with
\be  
F(Q^2) = \int_0^\infty \frac{\rho_F(s) ds}{s+Q^2}
\ee
and $\rho_F(s)$ a continuous spectral density.  
Therefore $F(Q^2)$ is the primary function of 
\[
-\Pi_{mq}(Q^2)/Q^2=\Pi_{mq}(q^2)/q^2\, .
\]
It reads
\be
F(Q^2) = -\int \frac{dQ^2}{Q^2}\Pi_{mq}(Q^2)= \int dL \; \Pi_{mq}(Q^2)
\quad L=\ln \frac{\mu^2}{Q^2} \, .
\ee
The result for $F(Q^2)$ is 
\be 
F(Q^2) = L + (\frac{7}{3} L + L^2)a +(15.757 L + 7.11 L^2 +1.417 L^3) a^2
+ ...
\ee
For the discontinuity across the cut defined by 
\be
\rho_F(s)= \frac{1}{2\pi i} \left(F(-s-i0)-F(-s+i0)\right)
\ee
one obtains
\ba
\label{roF}
\rho_F(s) &=& 1 +(\frac{7}{3}+2 l) a + (1.77 +14.22 l +4.25 l^2)a^2 + \\
      && (-207.04 + k_{q2} + 62.21 l + 54.52 l^2 + 9.21 l^3 ) a^3 + ...\nonumber
\ea
where we have already substituted the effective coupling $a$
for $\al_s$. Eq.~(\ref{roF}) has the standard form 
of the spectral density for needed for the comparison 
with the massless and the $g$ cases. 
In order to get rid of the constants the new mass parameter $m_q$
for the amplitude $\Pi_{mq}(Q^2)$
is defined in analogy with
the $g$-case such that
\be
m_q^2 =  m_s^2  \rho_F(M^2_\tau)
\ee
with $m_q^2 =m_q^2(M^2_\tau)$ and  
$m_s^2= m_s^2(M^2_\tau)$.
Then one explicitly has
\be
\label{mfms}
m_q^2 = m_s^2 (1 + \frac{7}{3} a + 1.77 a^2 + (-207.044 +k_{q2})a^3 
+ (-1335.5-2.33 k_3 - 4.92 k_{q2} + k_{q3}) a^4+...)  
\ee
Note that the order $a^4$ term contains not only 
the unknown coefficient 
$k_{q2}$ but also the higher order coefficient $k_{q3}$ 
which makes
the contribution of the $a^4$ term completely arbitrary. 
The definition of this new mass $m_q^2$ in terms of 
the $\MSsch$-scheme mass 
eq.~(\ref{mfms}) has
an accuracy of about 2\% if only second order corrections are used. 
The PT series is only known to third order 
-- already the fourth term contains the unknown
coefficient $k_{q2}$. 
Expressed through the $\MSsch$-scheme coupling constant $\al_s$
eq.~(\ref{mfms})
reads 
\be
\label{mfmsMS}
m_q^2=m_s^2\left(1+\frac{7}{3} \alsb + 5.60 \alsb^2 + 
        ( - 225.22 + k_{q2}) \alsb^3 + ...\right)
\ee
With this new mass $m_q^2$ a new spectral density
$\rho_q(s)$ can be defined in analogy to
eq.~(\ref{defofmg})
\be
m_s^2 \rho_F(s) = m_q^2 \rho_q(s)
\ee
that leads to the expansion 
\ba 
\label{densF}
\rho_q(s) &=& 1 + 2 l a  + (9.55 l +4.25 l^2 )a^2 + (36.36 l +44.6 l^2
+9.21 l^3)a^3 + \nonumber \\
&& ((-1141  -2 k_3 +6.75 k_{q2})l  +253.6 l^2 + 154.96 l^3 +20.14
l^4)a^4 + O(a^5)\, .
\ea
Note that the series (\ref{densF})
contains only logarithms and no constants.
The coefficients of the spectral density $\rho_q(s)$ from eq.~(\ref{densF}) 
are close to those of the $g$-part mass correction 
$\rho_g(s)$ eq.~(\ref{densg}) 
up to the second order. The coefficients of the highest powers of
logarithms are simply equal while the lower powers are different
because 
of different coefficients in the corresponding $D$-functions 
eq.~(\ref{fullinf}). The coefficient of the third order
term of eq.~(\ref{densF}) for $\rho_q(s)$
is seven 
times larger than the corresponding coefficient from eq.~(\ref{densg}) 
for $\rho_g(s)$ for the lowest power
of logarithm which dominates the 
behavior of the higher moments.   
The coefficients of the spectral density $\rho_q(s)$ are in general 
larger
than the 
coefficients of the spectral density $\rho(s)$ in the massless case
eq.~(\ref{densold}). 
Thus, the behavior of the spectral densities 
($\rho,\rho_g,\rho_q$) 
allows one to immediately conclude
about the convergence of the corresponding moments
for all three independent contributions up to order $m_s^2$ 
in the strange decays.

The standard moments of $\rho_q(s)$ as in 
eq.~(\ref{intmom},\ref{intmomg})
do not however coincide with the physical moments of 
the $q$-amplitude.
Indeed, the physical $q$-moments 
$M_q^{ph}(n)$
are defined through the contour integration in the following way
\be 
\label{defofqmom}
\frac{i m_s^2}{2 \pi}\oint \frac{\Pi_{mq}(q^2)}{q^2}
\left(\frac{q^2}{\mts}\right)^n dq^2 = m_q^2 M_q^{ph}(n),
\ee 
where the effective mass $m_q^2$
is used for normalization.
The moments eq.~(\ref{defofqmom}) can be evaluated through the 
function $\rho_q(s)$ directly.
The zero order moment turns out to be
\be
\label{qmom0}
M_q^{ph}(0) = -\rho_q(M_{\tau}^2) =  -1 \, .
\ee
The reason for this is that the integration with $n=0$ in 
eq.~(\ref{defofqmom}) picks up exactly those
contributions that eventually are absorbed into the strange quark
mass redefinition. 
Higher order physical moments 
are related to the 
standard moments of $\rho_q(s)$ via
\be
\label{qmom1}
M_q^{ph}(n)|_{n >0} = n\int_0^1 \rho_q(s)s^{n-1}ds - 1 \equiv 
M_q(n-1)-1 \, .
\ee
They
contain no 
parton model contribution.
The moments $M_q(n)$ are 
the standard objects defined in eq.~(\ref{intmom})
and the whole 
analysis of ref.~\cite{one} is applicable with $\rho_q(s)$ to be compared with
$\rho_g(s)$ and $\rho(s)$. 
Numerical values for the moments are
\ba
\label{qnmom}
M_q(0) &=& 1 + 2 a + 18.1 a^2 + 180.8 a^3 
+ (779.4 - 2 k_3 + 6.75 k_{q2}) a^4 \nonumber\\ 
M_q(1) &=& 1 + 1a + 6.90 a^2 + 47.39  a^3 
+ (-297.3 -k_3 + 3.375 k_{q2}) a^4 \nonumber \\
M_q(2) &=& 1 + \frac{2}{3}a + 4.13 a^2 + 24.08 a^3 
+ (-283.6 - 0.67 k_3 + 2.25 k_{q2})a^4 \nonumber\\
M_q(3) &=& 1 + \frac{1}{2} a + 2.92 a^2 + 15.53 a^3 
+ ( -237.13 - 0.5 k_3 + 1.69 k_{q2}) a^4 \nonumber \\
M_q(4)&=& 1 + \frac{2}{5}a + 2.25 a^2 + 11.28 a^3 
+ ( -199.699 - 0.4 k_3 + 1.35 k_{q2})a^4 \nonumber \\
&& \vdots  \nonumber\\
M_q(100)&=&1+\frac{1}{50} a+0.095a^2+0.37a^3+(-11.25 - 0.020 k_3 
+ 0.067 k_{q2})a^4 \, .
\ea 
The use of $\rho_q(s)$ is a universal way of describing the moments in FOPT
and easy for comparison with the massless and the $g$ parts. 
The physical $q$-moments $M_q^{ph}(n)$
are related to it by eq.~(\ref{qmom0})
and (\ref{qmom1}).
They do not, however, change the pattern of PT convergence. 
The moment $M_q^{ph}(1)$ related to $M_q(0)$
shows asymptotic growth in the third order already. 
As expected higher
moments show 
a better behavior because the low energy region of the integration
is
suppressed.

\section{Order $m_s^2$ corrections to $\tau$-lepton decay observables}
After introducing the technique for analyzing the moments 
at order $m_s^2$ we now apply 
it to an analysis of physical observables. 

\subsection{Total decay rate}
The $\tau$ decay width is given by a specific linear
combination of moments. The weight function contains
the overall factor $(1-s)^2$ which impairs
the convergence of the total decay rate observable.
The $(1-s)^2$ factor enhances the infra-red region of integration, i.e. 
the relative magnitude
of the contributions of logarithms $\ln(M_\tau^2/s)$
at small energy.
The concrete shape of the weight function with 
the weight factor $(1-s)^2$ is the main source of slow convergence
of the $m_s^2$ correction to the rate
\be \label{decayrate}
R_{m\tau} = \frac{i}{2 \pi} \oint 2 \left( 1- \frac{q^2}{\mts} \right)^2 
3\left(\frac{m_s^2 \Pi_{mq}(q^2)}{q^2} - \frac{m_s^2}{\mts}
\Pi_{mg}(q^2)  \right)\frac{dq^2}{\mts} \, .
\ee  
We retain the different mass definitions 
for the two invariant functions in order to explore the structure of 
the PT series and to check on its convergence.

The result for the mass correction of the total decay rate reads
\ba \label{decrthroumom}
R_{m\tau}&=&6\frac{m_q^2}{\mts}(M_q^{ph}(0)-2M_q^{ph}(1)+M_q^{ph}(2)) 
-6\frac{m_g^2}{\mts}(M_g(0)-M_g(1)+\frac{1}{3}M_g(2)) \nonumber \\
&=&-6 \frac{m_q^2}{\mts}(2 M_q(0)-M_q(1)) 
-6\frac{m_g^2}{\mts}( M_g(0)-M_g(1)+\frac{1}{3}M_g(2))   \, .
\ea
It is expressed through the effective mass parameters $m^2_{g,q}$
and the moments introduced earlier.
In the parton model approximation all the moments are normalized to
unity
which makes a glance analysis of eq.~(\ref{decrthroumom}) easy.
The convergence pattern of all the moments has been obtained 
already before.
Numerically one has
\ba 
\label{numdecayrate}
R_{m\tau} &=&-6 \frac{m_q^2}{\mts} \left( 1 + 3  a +  29.21 a^2 + 314.3 a^3 
+ (1856.1 - 3.0k_3 + 10.13 k_{q2}) a^4 \right) \nonumber \\
&&-2\frac{m_g^2}{\mts}
\left( 1 + 3.67 a + 34.84 a^2 + 337.3 a^3 + (3745.2-3.67 k_3) a^4 \right)
\ea
The reason for the bad convergence of (\ref{numdecayrate}) is the
contribution of the low moments
$M_q(0)$, $M_q(1)$ and $M_g(0)$, $M_g(1)$ 
to the mass correction of the total decay rate.
Both series in eq.~(\ref{numdecayrate}) converge only marginally
calling for the resummation of the series.
The total contribution is dominated by the $q$-part
which has a three times bigger coefficient in the leading term.  

Inserting the expressions for $m_q^2$, $m_g^2$, $a$ in terms of the
$\MSsch$-scheme parameters 
$m_s^2$ and $\alpha_s$ 
into eq.~(\ref{numdecayrate}) one obtains the standard result
\be 
R_{m\tau} = - 8 \frac{m_s^2}{\mts} \left( 1 + 5.33  \als + 46.0  \alsb^2 +
(283.55 + 0.75 k_{q2}) \alsb^3 + ...\right) \, .
\ee

\subsection{The '1+0' Method}
In ref.~\cite{exp} the numerical value for the 
strange quark mass in $\MSsch$-scheme has been extracted 
with the '1+0' method
which uses the representation of the total decay rate as a
sum of $(L+T)$ and $L$ contributions
(compare with eq.~(\ref{int-TL}))
\be \label{LT}
R_\tau = \frac{i}{2\pi}\oint 2 \left(1-\frac{q^2}{\mts} \right)^2 
\left\{ \left( 1 + 2\frac{q^2}{\mts} \right) \Pi_{(L+T)}(q^2) -2
\frac{q^2}{\mts} \Pi_L(q^2) \right\} \frac{dq^2}{\mts}\, ,
\ee

\be
\Pi_{(L+T)}(q^2) = \Pi_q(q^2) \, .
\ee
It is assumed 
that in the $\MSsch$-scheme the series for the $(L+T)$ part
of the $m_s^2$ correction in eq.~(\ref{LT}) 
converges well \cite{exp}.
The convergence is not impressive in FOPT 
but the numbers given for contour-improved FOPT in
ref.~\cite{exp} show fast convergence.
The quantity of interest is now 
\ba 
\label{Rtlt}
R_{m \tau}^{L+T}&=&\frac{i}{2\pi} 
\oint 2\left( 1- \frac{q^2}{\mts}\right)^2 
\left(1 + 2\frac{q^2}{\mts} \right)3\frac{\Pi_{mq}(q^2)}{q^2}
\frac{dq^2}{\mts}\nonumber \\
&=&6\frac{m_q^2}{\mts}(M_q^{ph}(0)-3M_q^{ph}(2)+2M_q^{ph}(3))\nonumber \\
&=&-6\frac{m_q^2}{\mts}(3M_q(1)-2M_q(2)).
\ea
If we look at the moments which are contained in $R_{m\tau}^{L+T}$
(\ref{Rtlt}) it is natural to expect 
a convergence behavior better than in the total decay rate because no zero
order moments are needed 
to construct $R_{m\tau}^{L+T}$. 
This is an invariant (scheme independent) reason for better
convergence: the particular combination $R_{m\tau}^{L+T}$ receives
smaller IR contribution from integration along the cut and therefore
is better computable in PT. 
Still the convergence is rather slow.  
Numerical results for the $(L+T)$ part are
\[
R_{m \tau}^{L+T}|_{q-scheme}=
\]
\be
- 6 \frac{m^2_q}{\mts} \left( 1 + 1.67 a + 12.448 a^2 + 94.01 a^3 
+ (-324.629 - 1.67 k_3 + 5.625 k_{q2})a^4 + \ldots\right)\, .
\ee
The convergence persists and the last term is still smaller 
than the third for the standard values of  $25<k_3<100$
and $0<k_{q2}<160$. The total contribution of first three terms 
is 0.45 which is a reasonable change of the leading order 
term due to PT correction.
In the $\MSsch$-scheme this becomes 
\ba
R_{m \tau}^{L+T}|_{\MSsch-scheme} &=&
-6 \frac{m_s^2}{\mts} \left( 1+ 4.0 \alsb + 24.67 \alsb^2 + (-62.77 +
k_{q2})\alsb^3  + \nonumber \right. \\ 
 && \left. (-3110 + 7.29 k_{q2} + k_{q3})\alsb^4 +... \right) 
\ea
The total contribution of the first two terms 
amounts to 0.65. The change in the leading order prediction
is even larger than the total change in 
the effective $q$-scheme where one more term of PT expansion
is available.
The only advantage of the '1+0` amplitude from PT point of view is the 
absence of the zero order moment $M_q(0)$ which 
is the most divergent one.
Still, the moment $M_q(1)$ from eq.~(\ref{qnmom})
which contributes to the 
rate expression eq.~(\ref{Rtlt})
is bad enough to prevent a fast
convergence.
The longitudinal part can be expressed in terms of the 
moments and convergence is
rather bad
\ba 
\label{lll}
R_{m \tau}^{L}&=&\frac{i}{2\pi}\oint 2\left(1-\frac{q^2}{\mts}\right)^2 
(-2)\frac{q^2}{\mts} \Pi_L(q^2) \frac{dq^2}{\mts} \nonumber \\
&=&-12\frac{m_q^2}{\mts} ( M_q(0) - 2 M_q(1) + M_q(2))
-6\frac{m_g^2}{\mts} (M_g(0) - M_g(1) + \frac{1}{3}M_g(2))
\ea
because again the lowest order moments enter.
The leading order parton model term in the $q$-part vanishes.
Numerical values are
\ba
\label{ll}
R_{m \tau}^{L}=&-&2\frac{m_g^2}{\mts}\left(1+3.67 a + 34.8 a^2 + 337.3 a^3 
+ (3745.2 - 3.67 k_3)a^4 \right) \nonumber \\  
&-&8\frac{m_q^2}{\mts} \left( 0+ a + 12.57 a^2 + 165.2 a^3 
+ (1635.6 - k_3 + 3.375 k_{q2})a^4 \right)
\ea
from which one can perceive the reason for the bad PT structure.
While the convergence of the $g$-part is rather standard, the admixture of
$q$-contribution
without the leading term is enhanced by relative factor of four 
which makes the sum in eq.~(\ref{ll}) completely uninterpretable.  

Note that we use $m_{g,q}$ mass parameters for $g,q$ parts of the
correlator as internal mass scales for $m_s^2$ corrections.
One can introduce another set of parameters $m_{T,L}$
related to the definite spin decomposition of the correlator 
eq.~(\ref{spindecomp}). As one can see from eq. (\ref{physdec})
$m_T=m_g$ because the $\Pi_T(q^2)$ amplitude is proportional to 
$\Pi_g(q^2)$. The effective mass $m_L^2$ is obtained from th longitudinal
part $\Pi_L(q^2)$. The expression of $m_T^2$ through 
the $\MSsch$-scheme mass is reasonable (eq.~(\ref{mg})) 
while the corresponding relation 
for $m_L$ is much wilder.

\begin{samepage}
\ba
m_L^2 &=& m_s^2 \left( 1+ 5.67 \als + 31.9 \alsb^2 + 89.2 \alsb^3 + \right. \nonumber \\
      &&   \left. (- 5180 + k_{g3} + 17.5 k_{q2}) \alsb^4 + ... \right)
\ea
\end{samepage}
The different structures of PT serieses for the parameters $m_T^2$ ans $m_L^2$ 
can be interpreted as a result of the difference of full QCD interaction in spin 1 
and spin 0 channels.
The stronger and earlier 
breakdown of PT behavior in the spin 0 channels
can be related to the contribution of 
non-perturbative effects (instantons) which is absent in 
spin 1 channel.

To summarize,
the convergence of the PT series for the $m_s^2$ corrections
for the most natural and precisely measured physical observables
is always slow and almost marginal.
This is because the IR region of integration is numerically important
for the given value of the coupling constant and the order of PT. 
Any apparent fast
convergence is the result of either a specific linear combination of
moments
or a particular scheme choice.
The former case is, however, not realized for physical
observables of interest measurable in experiment.
This would imply that a resummation of the series 
is necessary for a sound interpretation of
the theoretical formulas for decay rates.
 
\section{Conclusions}
We have analyzed the asymptotic structure
of PT series for $m_s^2$ corrections.
Using the standard estimate of the accuracy of an asymptotic
series we have found 
that the theoretical precision 
in the perturbative description of Cabibbo supressed
$\tau$-lepton decays 
is already limited by the
asymptotic growth of the coefficients in fourth order of perturbation
theory. 
This is a scheme invariant
statement. 
The accuracy of the perturbative expansion for the coefficient
functions of the $m_s^2$ corrections in Cabibbo suppressed 
channels are 15\%-20\% at best.
Therefore the extraction of the numerical value for the strange quark
mass
from the $m_s^2$ corrections to the $\tau$-decay rate into strange hadrons
is limited by the precision of the coefficient functions.
Better theoretical accuracy can be obtained by using observables which contain 
higher order moments but 
the experimental accuracy of them is not good enough at present.
From phenomenological point of view the analysis of the $m_s^2$ corrections
differ from the massless case. While in the latter the low order moments
can be excluded by substituting experimantal results for them (from $e^+e^-$ 
annihilation, for instance), the coefficient functions of 
$m_s^2$ corrections have no immediate physical meaning and cannot be traded for 
in this manner. 

The introduction of two natural mass parameters allows to describe massless,
$q$ and $g$ parts of the correlator 
up to order $a^4$ with only two unknown parameters 
$k_3$ and $k_{q2}$ instead of four in the $\MSsch$-scheme. 
The existence of two different mass scales is physically motivated
by the difference of interaction in spin 1 and spin 0 channels. 
Still for both independent $m_s^2$ 
corrections ($g$ and $q$ part) the convergence of low order moments
is slow. The contribution of IR region is large and these quantities
have limited precision when evaluated within PT.
It can not be improved by a particular change of the scheme or 
taking particular linear combinations. 
Only those moments converge well
where the IR contribution is suppressed.
The renormalization group improved QCD parameters $a(s)$, $m_s(s)$ 
run too fast to give
precise PT series for a set of $\tau$ observables containing low moments with 
large IR contribution. 
If the running would be slower -- 
coefficients of $\beta$- and $\gamma$-functions would
be smaller -- then one could meet the high standards of experimental
precision for low moments with FOPT theoretical formulas.
Observables with higher moments are described well within 
FOPT but experimental accuracy is not yet sufficient for 
precise comparison. 

Therefore for
precise comparison of theory with experiment  
some procedure of resummation
is required \cite{Pivtau,renRS,groote}.
And, in fact, 
one can perform
resummation 
only for lowest order moments to keep things close 
to the standard FOPT. Resumation, however, introduces additional 
(to the standard renormalization group freedom \cite{RG})
arbitrariness, in particular, it interferes with nonperturbative power
corrections that makes separation of contributions 
within the OPE not unique.

This problem is beyond the scope of the present analysis.

\subsection*{Acknowledgments}
The work is supported in part by the Volkswagen 
Foundation under contract No.~I/73611 and 
by the Russian Fund for Basic Research under contracts
Nos.~97-02-17065 and 99-01-00091. 
A.A.~Pivovarov is an Alexander von Humboldt fellow.

\newpage
\section*{Appendix}
The input for all calculations are the coefficients of the correlator, the
beta and the gamma function which are available up to 
four loop calculations in present.
In the case of $\Pi_q(Q^2)$ only corrections up to $\al_s^2$ 
are known because the 
constant term of the correlator is not yet calculable. 
All coefficients are given 
in the $\MSsch$-scheme (see, e.g. \cite{phys_report}).
Below $\zeta(z)$ is the Riemann $\zeta$-function.

The coefficients of the correlator are \cite{eek2,eek2c}
\ba
k_1 &=&  \frac{299}{24}- 9 \zeta(3), \nonumber \\
k_2 &=& \frac{58057}{288} - \frac{779}{4}\zeta(3) +\frac{75}{2}\zeta(5),
\nonumber \\
k_{q1} &=& \frac{13981}{432} +
\frac{323}{54}\zeta(3)-\frac{520}{27}\zeta(5), 
                             \nonumber \\
k_{g1} &=& \frac{4591}{144} - \frac{35}{2} \zeta(3), \nonumber \\
k_{g2} &=& \frac{1967833}{5184} - \frac{\pi^4}{36}
-\frac{11795}{24}\zeta(3) +  \frac{33475}{108} \zeta(5) \, .
          \nonumber
\ea
Beta function coefficients \cite{beta4}
\ba
\beta_0 &=& \frac{9}{4}, \;\; \beta_1 = 4, \;\; \beta_2 =
\frac{3863}{384}, \nonumber \\
\beta_3 &=& \frac{140599}{4608} + \frac{445}{32} \zeta(3) \, .
\nonumber
\ea
Gamma function coefficients \cite{gamma3c,gamma3v}
\ba
\gamma_0 &=& 1, \;\; \gamma_1 = \frac{91}{24}, \;\; \gamma_2 =
\frac{8885}{578}- \frac{5}{2} \zeta(3), \nonumber \\
\gamma_3 &=& \frac{2977517}{41472} + \frac{3 \pi^4}{32}-
\frac{9295}{432}\zeta(3)-\frac{125}{12}\zeta(5) \, . 
\nonumber
\ea


\begin{thebibliography}{99}
\bibitem{one}J.G.~K\"orner, F.Krajewski, A.A.~Pivovarov,
MZ-TH/99-37, \\Eur.Phys.J. C, to be published;
hep-ph/9909225.
\bibitem{exp}ALEPH collaboration, Z. Phys. C76(1997)15,
Eur.Phys.J. C4(1998)409,\\ 
CERN-EP/99-026.
\bibitem{PDG}Particle Data Group, Review of Particle Properties,
Eur.Phys.J. C3(1998)1. 
\bibitem{msNPB}K.G. Chetyrkin, J.H. K\"uhn, A.A. Pivovarov,
Nucl. Phys. B533 (1998) 473.
\bibitem{SchTra84} 
K.Schilcher, M.D. Tran, Phys.\ Rev.\ D 29 (1984) 570.
\bibitem{Bra88}
E.Braaten, Phys.\ Rev.\ Lett.\ 53 (1988) 1606.
\bibitem{Bra89} 
E.Braaten, Phys.\ Rev.\ D 39 (1989) 1458.
\bibitem{NarPic88}
S.Narison, A.Pich,  Phys.\  Lett.\ B 211 (1988) 183.
\bibitem{BraNarPic92} 
E.Braaten, S.Narison, A.Pich, Nucl.\ Phys.\ B 373 (1992) 581.
\bibitem{chet93}K.G. Chetyrkin, Phys.\ Lett.\ B 307 (1993) 169.
\bibitem{SVZ}M.A.Shifman, A.I.Vainshtein, V.I.Zakharov, Nucl.Phys.
B147(1979)385
\bibitem{cont}
C.Bernard, A.Duncan, J.LoSecco, S.Weinberg,
Phys.Rev. D12 (1975) 792;\\ 
E.Poggio, H.Quinn, S.Weinberg,
Phys.Rev. D13 (1976) 1958.
\bibitem{cont1}R.Shankar, Phys.Rev. D15 (1977) 755.
\bibitem{cont2}K.G.Chetyrkin, N.V.~Krasnikov, A.N.Tavkhelidze,
Phys.Lett. 76B (1978) 83.
\bibitem{Pivtau}A.A. Pivovarov, Sov. J. Nucl. Phys. 54(1991) 676;
Z.Phys. C53(1992)461; Nuovo Cim. 105A(1992)813.
\bibitem{effsch}
G. Grunberg, Phys.Lett. 95B(1980)70, Erratum-ibid. 110B(1982)501 
\bibitem{ksch}
N.V.Krasnikov, Nucl.Phys. B192 (1981) 497
\bibitem{kksch}A.L. Kataev, N.V.Krasnikov, A.A.Pivovarov,
Phys.Lett. 107B (1981) 115;\\ 
Nucl.Phys. B198 (1982) 508 
\bibitem{effDh}A.Dhar and V.Gupta, Phys.Rev. D29(1984)2822 
\bibitem{prl}S.Groote, J.G.K\"orner, A.A.Pivovarov, K.Schilcher,
Phys.Rev.Lett. 79 (1997) 2763
\bibitem{renRS}N.V.~Krasnikov, A.A.~Pivovarov, Mod.Phys.Lett. 
A11(1996)835.
\bibitem{brodsky1}
S.J. Brodsky, J.R. Pelaez, N. Toumbas,
Phys.Rev.D60(1999)037501
\bibitem{brodsky}J.R. Pelaez, S.J. Brodsky, N. Toumbas, 
SLAC-PUB-8147, May 1999.\\
Talk given at 34th Rencontres de Moriond: 
QCD and Hadronic Interactions, \\
Les Arcs, France, 20-27 Mar 1999; hep-ph/9905435 
\bibitem{groote}S. Groote, J.G. K\"orner, A.A. Pivovarov,\\
Phys. Lett. B407(1997)66,
Mod.Phys.Lett. A13 (1998) 637
\bibitem{RG}N.N.Bogoliubov and D.V.Shirkov, Quantum fields (Benjamin,
1983).
\bibitem{phys_report}K.G.~Chetyrkin, J.H.~K\"uhn and A.~Kwiatkowski, 
Phys.~Rep.\ { 277} (1996) 189   
\bibitem{eek2}S.G.Gorishny, A.L.Kataev and S.A.Larin,
Phys.Lett. B259(1991)144;\\
L.R.Surguladze and M.A.Samuel, Phys.Rev.Lett. 66(1991)560, 2416(E),\\
Phys.Rev. D44(1991)1602.
\bibitem{eek2c}K.G.~Chetyrkin,  Phys.Lett. B391(1997)402 
\bibitem{beta4}T. van Ritbergen, J.A.M.Vermaseren and S.A.Larin,  
Phys.Lett. B400(1997)379
\bibitem{gamma3c}K.G.~Chetyrkin,  Phys.Lett. B404(1997)161 
\bibitem{gamma3v}T. van Ritbergen, J.A.M.Vermaseren and S.A.Larin,  
Phys.Lett. B405(1997)327
\end{thebibliography}
\end{document}